\documentclass[twocolumn,trackchanges]{aastex61}
\usepackage{booktabs}
\usepackage{color}
\usepackage{soul}
\usepackage{ulem}
\usepackage{bm}
\usepackage{amsmath}

\hypersetup{linkcolor=red,citecolor=green,filecolor=cyan,urlcolor=magenta}

\submitjournal{ApJS}

\shortauthors{Wang et al.}

\begin{document}

\title{B\lowercase{enchmarking} A\lowercase{tomic} D\lowercase{ata} \lowercase{for} A\lowercase{strophysics}: B\lowercase{e-like} I\lowercase{ons} \lowercase{between} B~II \lowercase{and} N\lowercase{e}~VII}

\author{Kai Wang}
\affil{Hebei Key Lab of Optic-electronic Information and Materials, The College of Physics Science and Technology, Hebei University, Baoding 071002, China;}
\affil{Group for Materials Science and Applied Mathematics, Malm\"o University, SE-20506, Malm\"o, Sweden; {\color{blue} per.jonsson@mah.se}\\}
\affil{Shanghai EBIT Lab, Key Laboratory of Nuclear Physics and Ion-beam Application, Institute of Modern Physics, Department of Nuclear Science and Technology, Fudan University, Shanghai 200433, China; {\color{blue} chychen@fudan.edu.cn}}

\author{Zhan Bin Chen}
\affiliation{College of Science,  Hunan University of Technology, Zhuzhou 412000, China}
\author{Chun Yu Zhang}
\affil{Shanghai EBIT Lab, Key Laboratory of Nuclear Physics and Ion-beam Application, Institute of Modern Physics, Department of Nuclear Science and Technology, Fudan University, Shanghai 200433, China; {\color{blue} chychen@fudan.edu.cn}}
\author{Ran Si}
\affil{Shanghai EBIT Lab, Key Laboratory of Nuclear Physics and Ion-beam Application, Institute of Modern Physics, Department of Nuclear Science and Technology, Fudan University, Shanghai 200433, China; {\color{blue} chychen@fudan.edu.cn}}
\author{Per J\"onsson}
\affil{Group for Materials Science and Applied Mathematics, Malm\"o University, SE-20506, Malm\"o, Sweden; {\color{blue} per.jonsson@mah.se}\\}
\author{Henrik Hartman}
\affiliation{Group for Materials Science and Applied Mathematics, Malm\"o University, SE-20506, Malm\"o, Sweden; {\color{blue} per.jonsson@mah.se}\\}
\author{Ming Feng Gu}
\affiliation{Space Science Laboratory, University of California, Berkeley, CA 94720, USA}


\author{Chong Yang Chen}
\affil{Shanghai EBIT Lab, Key Laboratory of Nuclear Physics and Ion-beam Application, Institute of Modern Physics, Department of Nuclear Science and Technology, Fudan University, Shanghai 200433, China; {\color{blue} chychen@fudan.edu.cn}}

\author{Jun Yan}
\affil{Institute of Applied Physics and Computational Mathematics, Beijing 100088, China;  {\color{blue} yan$_{-}$jun@iapcm.ac.cn} }
\affil{Center for Applied Physics and Technology, Peking University, Beijing 100871, China;}
\affil{Collaborative Innovation Center of IFSA (CICIFSA), Shanghai Jiao Tong University, Shanghai 200240, China}



\begin{abstract}
Large-scale self-consistent multiconfiguration Dirac--Hartree--Fock  and relativistic configuration interaction  calculations  are reported for the $n \leq 6$ levels in Be-like ions from \ion{B}{2} to \ion{Ne}{7}. Effects from electron correlation are taken into account by means of large expansions in terms of a basis of configuration state functions, and a complete and accurate data set of excitation energies, lifetimes, wavelengths, and electric dipole, magnetic dipole, electric quadrupole, and magnetic quadrupole line strengths, transition rates, and oscillator strengths for these levels is provided for each ion. Comparisons are made with available experimental and theoretical results. The uncertainty of excitation energies is assessed to be 0.01\% on average, which makes it possible to find and rule out misidentifications and aid new line identifications involving high-lying levels in astrophysical spectra. The complete data set is also useful for modeling and diagnosing astrophysical plasmas.

\end{abstract}




\section{Introduction} \label{sec:intro}
A wealth of astrophysical spectra has been obtained from different missions, such as Chandra, X-ray Multi-Mirror Mission (XMM-Newton), Hinode, and Hard X-ray Modulation Telescope (HXMT). 
The analysis of these expensively acquired spectra is often not limited by the capabilities of the spectrometers themselves, but by the lack of accurate reference data in the atomic line database~\citep{Bigot.2006.V372.p609,Kallman.2007.V79.p79,Ruffoni.2013.V779.p17}.
To solve this dilemma, we have computed highly accurate energy and radiative transition data for L-shell atomic ions~\citep{Jonsson.2013.V559.p100,Ekman.2014.V564.p24,Wang.2014.V215.p26,Wang.2015.V218.p16,Wang.2016.V223.p3,Wang.2016.V226.p14,Wang.2017.V229.p37,Radziute.2015.V582.p61,Si.2016.V227.p16}. This work reports accurate atomic data resulting from our systematic calculations for beryllium-like ions from \ion{B}{2} to \ion{Ne}{7}.

These ions have been observed in spectra from different astrophysical  objects, such as the sun~\citep{Curdt.1997.V126.p281,Curdt.2001.V375.p591,Curdt.2004.V427.p1045,Feldman.1997.V113.p195,DelZanna.2011.V528.p139,Ko.2002.V578.p979,Parenti.2005.V443.p679,Tian.2009.V505.p307,Thomas.1994.V91.p461}, white dwarfs~\citep{Raassen.2002.V389.p228,Werner.2004.V427.p685} , as well as nebular regions in RR Tel~\citep{Penston.1983.V202.p77}. The atomic lines do not only tell what elements are present in astrophysical objects and what are their relative abundances, but also reveal the physical conditions in the plasmas, such as density, temperature and radiation 
fields, along with important excitation mechanisms and photo-processes. For example, the $n=3 \rightarrow 2$ \ion{O}{5} lines identified by the Extreme Ultraviolet (EUV) Imaging Spectrometer of the Hinode are density sensitive and  can be used for density diagnostics of the solar corona~\citep{Landi.2009.V706.p1}. \ion{C}{3} lines have been observed in NGC 2440 and provided carbon abundance information for this planetary nebula~\citep{Rubin.2004.V605.p784}.  Due to the high cosmic abundance from these light elements, the lines often appear strong, increasing their value as probes of physical conditions.

Atomic data for low-lying levels $(1s^2) 2l^2$ and $2l 3l^{\prime}$ with $l \leq 1$ and  $l ^{\prime}\leq 2$ (the $n = 2,  3$ complexes)  of the ions from \ion{B}{2} to \ion{Ne}{7} have previously been obtained from different calculations~\citep{Safronova.1996.V53.p4036,Safronova.1997.V30.p2375,Safronova.1999.V32.p3527,Tachiev.1999.V32.p5805,Komasa.2002.V65.p42507,Galvez.2003.V118.p6858,Fischer.2004.V87.p1,Gu.2005.V89.p267,Ho.2006.V74.p22510,Cheng.2008.V77.p52504,Marques.2012.V86.p52521,Verdebout.2014.V100.p1111}.  For many objects, however, emission lines from higher-excitation lines are observed. In the spectrum of the active Seyfert Galaxy NGC~1068 (Brinkman) and the supergiant $\delta$ Orionis ~\citep{Raassen.2013.V550.p55}, lines from levels with $n>3$ in the light elements such as C, N, O are identified.

There is thus a need for corresponding atomic data for the $n > 3$ higher-lying levels to be provided. Our present study is improving existing data, and extending the included levels to $n \leq 6$ as is described below. 
The new data will extend the probes of physical conditions in different astrophysical objects, as well as of analyzing new observations from different space missions and laboratory experiments~\citep{Traebert.2014.V215.p6,Traebert.2014.V211.p14}.

The energies for the $2lnl^{\prime}~^1\!S$ ($n=2-6$) states and transition rates for the $2lnl^{\prime}~^1\!S~-~2s2p~^1\!P$ transition  were calculated by~\citet{Zhang.2005.V16.p951} for the beryllium isoelectronic sequence ($Z$ = 4--10) using a configuration interaction method. Excitation energies of the $2pns~^{1,3}\!P$  and $2pnd~^{1,3}\!P$  Rydberg states ($n$ = 3--60) in \ion{B}{2} were semi-empirically determined by~\citet{Sakho.2013.V99.p447} in the framework of the screening-constant-by-unit-nuclear-charge method.  ~\citet{Savukov.2006.V39.p2115} performed many-body perturbation theory (MBPT) calculations for a large number of energy levels in Be-like \ion{Ne}{7}. The AUTOSTRUCTURE calculations of~\citet{Menchero.2014.V566.p104} were performed for the $n > 3$ levels of Be-like ions with $Z \leq 10$. Data for the $n > 3$ states of \ion{N}{4} were provided by ~\citet{Aggarwal.2016.V461.p3997} using the GRASP0 code~\citep{Grant.1980.V21.p207}, and by~\citet{Menchero.2017.V50.p65203} using the B-spline box-based close-coupling method~\citep{Zatsarinny.2009.V180.p2041}.

Among the different studies involving the $n >3$ levels the one by ~\citet{Savukov.2006.V39.p2115} for \ion{Ne}{7} provides energy data of high accuracy. In comparison, although complete sets of data including transition rates are provided, the studies by~\citet{Menchero.2014.V566.p104},  for the ions with $Z =5 -10$, and by ~\citet{Aggarwal.2016.V461.p3997} and ~\citet{Menchero.2017.V50.p65203},  for \ion{N}{4}, are quite inaccurate because of limited configuration interaction effects included in the calculations. For example, the excitation energies of ~\cite{Menchero.2014.V566.p104}, ~\citet{Aggarwal.2016.V461.p3997}, and ~\citet{Menchero.2017.V50.p65203} for \ion{N}{4} depart from the experimental energies compiled in the Atomic Spectra Database (ASD) of the National Institute of Standards and Technology (NIST)~\citep{Kramida.2015.V.p}  by up to 5.8\%, 5.7\%, and 2.4\%, respectively.  
These uncertainties are too large for identification and deblending of  new observations from different space missions. 

To fill the gaps of lacking accurate atomic data for high-lying states of  the ions from \ion{B}{2} to \ion{Ne}{7},  in particular accurate transition rates, 
we performed large-scale multiconfiguration Dirac--Hartree--Fock (MCDHF) and relativistic configuration interaction (RCI) calculations using the latest version of the GRASP2K code~\citep{Jonsson.2013.V184.p2197}. The calculations provide a consistent and highly accurate data set of energy structure and radiative transition parameters involving also the $n >3$ levels for these low charged  Be-like ions. This work is a continuation of our previous calculations \citep{Wang.2015.V218.p16}, in which the corresponding  accurate results are reported for the ions in the range of nuclear charges $ 10 \le Z \le 36$. The excitation energies and lifetimes for the levels of the $2l^2$ (with  $l \leq 1$) and $2l nl^{\prime}$ (with  $n \leq 6$ and  $l ^{\prime}\leq n-1$) configurations, which are below the levels of  the $2s 7l$ configuration, and electric dipole (E1), magnetic dipole (M1), electric quadrupole (E2), and magnetic quadrupole (M2) transition rates among these states are calculated for each ion.  To assess the accuracy of the present MCDHF/RCI data, 
MBPT calculations for \ion{Ne}{7} are carried out using the Flexible Atomic Code (FAC)~\citep{Gu.2008.V86.p675}. The two sets of results are in excellent agreement. Compared with previous calculations of Be-like ions, the present work results in a significant extension of accurate energy and transition data for higher-lying states of the $n > 3$ configurations, which will greatly aid the analysis of new spectra from astrophysical sources, as well as  improving the assessment of blending for diagnostic lines of interest.

\section{Theory and Calculations}
\subsection{MCDHF}\label{Sec:MCDHF}
The MCDHF method has been described in detail by~\citet{Grant.2007.V.p} and~\citet{FroeseFischer.2016.V49.p182004}. The method is also outlined in our previous calculations~\citep{Jonsson.2013.V559.p100,Ekman.2014.V564.p24,Wang.2016.V223.p3,Wang.2016.V226.p14,Wang.2017.V229.p37,Wang.2017.V117-118.p174,Wang.2017.V117-118.p1,Radziute.2015.V582.p61,Si.2016.V227.p16,Chen.2017.V114.p61,Chen.2017.V113.p258}.  Based on the active space (AS) approach~\citep{Olsen.1988.V89.p2185,Sturesson.2007.V177.p539} for the generation of the configuration state function (CSF) expansions, separate  calculations are done for the even and odd parity states. 
For the even parity states, the CSF expansions are obtained by allowing single and double (SD) excitations from the multi-reference (MR) configurations $2s^2$, $2p^2$, $2s 3s$, $2s 3d$, $2p 3p$, $2s 4s$, $2s 4d$, $2p 4p$, $2p 4f$, $2s 5s$, $2s 5d$, $2s 5g$, $2p 5p$, $2p 5f$, $2s 6s$, $2s 6d$, $2s 6g$, $2p 6p$, $2p 6f$, and $2p 6h$ to an AS of orbitals. 
For the odd parity states, the CSF expansions are obtained by allowing SD excitations from the MR configurations $2s 2p$, $2s 3p$, $2p 3s$, $2p 3d$, $2s 4p$, $2s 4f$, $2p 4s$, $2p 4d$, $2s 5p$, $2s 5f$, $2p 5s$, $2p 5d$, $2p 5g$, $2s 6p$, $2s 6f$, $2s 6h$, $2p 6s$, $2p 6d$, and $2p 6g$ to an AS of orbitals. In the first step of the calculations, the AS is

${\rm AS_1} = \{7s, 7p, 7d, 7f, 7g, 7h, 7i\}$.
\\

Then, the AS is  increased in the following way:\\

${\rm AS_2 = AS_1}+\{8s, 8p, 8d, 8f, 8g, 8h, 8i, 8k\}$,
\\

${\rm AS_3 = AS_2}+\{9s, 9p, 9d, 9f, 9g, 9h, 9i, 9k\}$,
\\

${\rm AS_4 = AS_3}+\{10s, 10p, 10d, 10f, 10g, 10h, 10i, 10k\}$,
\\

${\rm AS_5 = AS_4}+\{11s, 11p, 11d, 11f, 11g, 11h, 11i, 11k\}$
\\

SD excitations from the $1s$ subshell to active sets with principal quantum numbers $n \leq 7$ are allowed, whereas at most one excitation from the $1s$ subshell is allowed to active sets with principal quantum numbers $8 \leq n  \leq 11$. The final model using the AS$_5$ active set contains 
777~325  even and 800~410  odd parity CSFs. The Breit interaction and leading QED effects (vacuum polarization and self-energy) are included in subsequent RCI calculations. 
All calculations were performed using the GRASP2K code~\citep{Jonsson.2007.V177.p597,Jonsson.2013.V184.p2197}.

\subsection{MBPT}
The MBPT method is explained in~\cite{Lindgren.1974.V7.p2441,Safronova.1996.V53.p4036,Vilkas.1999.V60.p2808,Gu.2005.V156.p105,Gu.2007.V169.p154}, and it has been implemented in the FAC package~\citep{Gu.2008.V86.p675}. 
The key feature of the MBPT method is the partitioning of the Hilbert space of the system into two subspaces, the model space $M$ and the orthogonal space $O$. The configuration interaction effects in the $M$ space are exactly considered, while the interaction between the space $M$ and $O$ is
taken into account with the second-order perturbation method. 
For the MBPT calculation, the model space \emph{M} contains the even and odd multi-reference configurations of the MCDHF/RCI method,
while the space \emph{O} contains all the possible configurations
that are generated by SD virtual excitations of the \emph{O} space.
For single and double excitations, the maximum \emph{n} values are 125 and 65, respectively, with a maximum $l$ value of 25. Just as for the MCDHF/RCI calculations, QED effects are also included.

\section{EVALUATION OF DATA}
\subsection{Energy Levels}
The excitation energies for the lowest 138 states  of the $2s^2, 2p^2$ and $2l nl^{\prime}$ (with  $n \leq 6$, with  $l \leq 1$ and $l ^{\prime}\leq n-1$) configurations with $Z=8-10$ from our MCDHF/RCI calculations are listed in Table~\ref{table.en}, along with the results for the lowest 100 states of these configurations with $Z=5-7$. All these levels are below the levels of the $2s 7l$ configurations.
In relativistic calculations the wave functions for the states are given as expansions over $jj$-coupled CSFs. This labeling system is rarely used in databases or by experimentalists. To overcome this labeling problem,  the method developed ~\citet{Gaigalas.2004.V157.p239,Gaigalas.2017.V5.p6} is used to transform wave functions from the $jj$-coupling to the $LSJ$-coupling scheme. In Table~\ref{table.en},  for each level numbered by a key $(\#)$, the $LSJ$ designation, the total angular momentum and parity $J^{\pi}$, and the radiative lifetime estimated from theoretical transition rates are also included. 
 
Among the six Be-like ions considered here, both experimental and theoretical energy data of the levels up to the $n=6$ configurations  are relatively complete 
for \ion{Ne}{7}. In what follows we will firstly access the accuracy of the MCDHF/RCI excitation energies for 
\ion{Ne}{7}, by comparing available results. In Table~\ref{table.as}, we present the MCDHF/RCI excitation energies of the 138 levels for \ion{Ne}{7}  as a function of the increasing active set (AS). For comparison, the compiled values from the NIST ASD~\citep{Kramida.2015.V.p} are given as well. 
The mean differences between the MCDHF/RCI and NIST excitation energies and the corresponding standard deviations are
$-868 \pm 1724$ cm$^{-1}$, $-439 \pm 625$ cm$^{-1}$, $-240 \pm 264 $ cm$^{-1}$, $-179  \pm 219$, and $-170  \pm 206$ cm$^{-1}$ for the
calculations based on $\rm AS_1$, $\rm AS_2$, $\rm AS_3$, $\rm AS_4$, and $\rm AS_5$, respectively. 
This implies that the MCDHF/RCI calculations are well converged with respect to an increasing size of the AS and that the accuracy cannot be further improved by extending the orbital set. Remaining energy differences are due to
higher order correlation effects not captured within the framework of SD excitation from the MR. For AS$_5$ the standard deviation given above corresponds to an average relative difference 
of $ -0.01~\% 
 \pm 0.02~\%$.

In Table~\ref{table.en.ne},  computed excitation energies of \ion{Ne}{7} from the present MCDHF/RCI and MBPT calculations and from the calculations by~\citet{Savukov.2006.V39.p2115} [hereafter referred to as MBPT2],\\ ~\citet{Tachiev.1999.V32.p5805,Fischer.2004.V87.p1} [MCHF/BP],  ~\citet{Safronova.1996.V53.p4036,Safronova.1997.V30.p2375} [MBPT3], and by ~\cite{Menchero.2014.V566.p104} [AUTOSTRUCTURE] are compared with the NIST compiled values. The present MBPT calculations used 
the same method and code as in~\citet{Wang.2015.V218.p16} and ~\citet{Gu.2005.V89.p267}. Due to these similarities, the latter results are not shown in Table~\ref{table.en.ne}.

Compared with the MBPT3 and MCHF/BP calculations, the present MCDHF/RCI and MBPT calculations, as well as the MBPT2 calculations, report data for higher states. Moreover, the latter three calculations show better agreement with the NIST experimental values.  The average differences with the standard deviations between the computed excitation energies and the NIST values for the $n \leq 3$ levels are $-129 \pm 174$ cm$^{-1}$ for MCDHF/RCI, $-134 \pm 281$ cm$^{-1}$ for MBPT, and $100 \pm  372$ cm$^{-1}$ for MBPT2. The corresponding values for MBPT3 and MCHF/BP are $-303 \pm 609$ cm$^{-1}$ and $468 \pm 211$ cm$^{-1}$, respectively.

Inspecting all the $n \leq 6$ results of the MCDHF/RCI, MBPT, and MBPT2 calculations for \ion{Ne}{7} more carefully, one can see that the accuracies of three different calculations, for which the relative average energy differences with the NIST values are $-0.01~\% \pm  0.02~\%$ (MCDHF/RCI), $-0.01~\% \pm  0.03~\%$ (MBPT), and $0.01~\% \pm  0.06~\%$ (MBPT2), respectively, are generally at the same level. 
The largest deviations with the NIST values are -776 cm$^{-1}$ for  $\#122 / 2s 6p~^{1}P_{1}$ of MCDHF/RCI, -1111 cm$^{-1}$ for $\#37 /2p 3p~^{1}D_{2}$ of MBPT, and 2102 cm$^{-1}$ $\#10/2p^{2}~^{1}S_{0}$ of MBPT2. In comparison, although providing complete sets of energy data along with collision strengths, the AUTOSTRUCTURE calculations~\citep{Menchero.2014.V566.p104},  show a much larger average difference with the NIST values, $-2229\pm 2625$ cm$^{-1}$, which is far greater than the requirement of analyses of new astrophysical spectra.

Experimental and theoretical energy data are available for many $n \leq 3$ levels  along the isoelectronic sequence from \ion{B}{2} to \ion{Ne}{7}. To further assess the accuracy of our computed energy values, a comparison with the NIST experimental values as well as with values from the MBPT3 and MCHF/BP calculations, is given in Table~\ref{table.en.n23}.  From the table it is clear that the present MCDHF/RCI calculations provide more theoretical data compared with the two previous calculations. The differences with the NIST experimental values are also much smaller for our calculations than those from the MBPT3 and MCHF/BP calculations. More specifically, excluding 
the levels $2p3p~^{3}S_{1}$ with $Z=5,7$, $2p3d~^{3}P_{0}^{\circ}$ with  $Z=5$, and  $2p3p~^{1}S_{0}$  with $Z=9$, for which the differences of theory from experiment are greater than 1000 cm$^{-1}$ (they will be further discussed below), the average differences with standard deviations, $\Delta E \pm \sigma$, with the NIST values are $-23 \pm 141$ cm$^{-1}$ for MCDHF/RCI, $-630 \pm 1854$ cm$^{-1}$ for MBPT3, and $256 \pm 173$ cm$^{-1}$ for MCHF/BP. It should be noted that the differences of the MBPT3 and NIST values decrease with increasing  $Z$ for the same level and that the large uncertainties for this method mainly come from the low-charge ions.

\begin{figure*}[ht!]
	\plotone{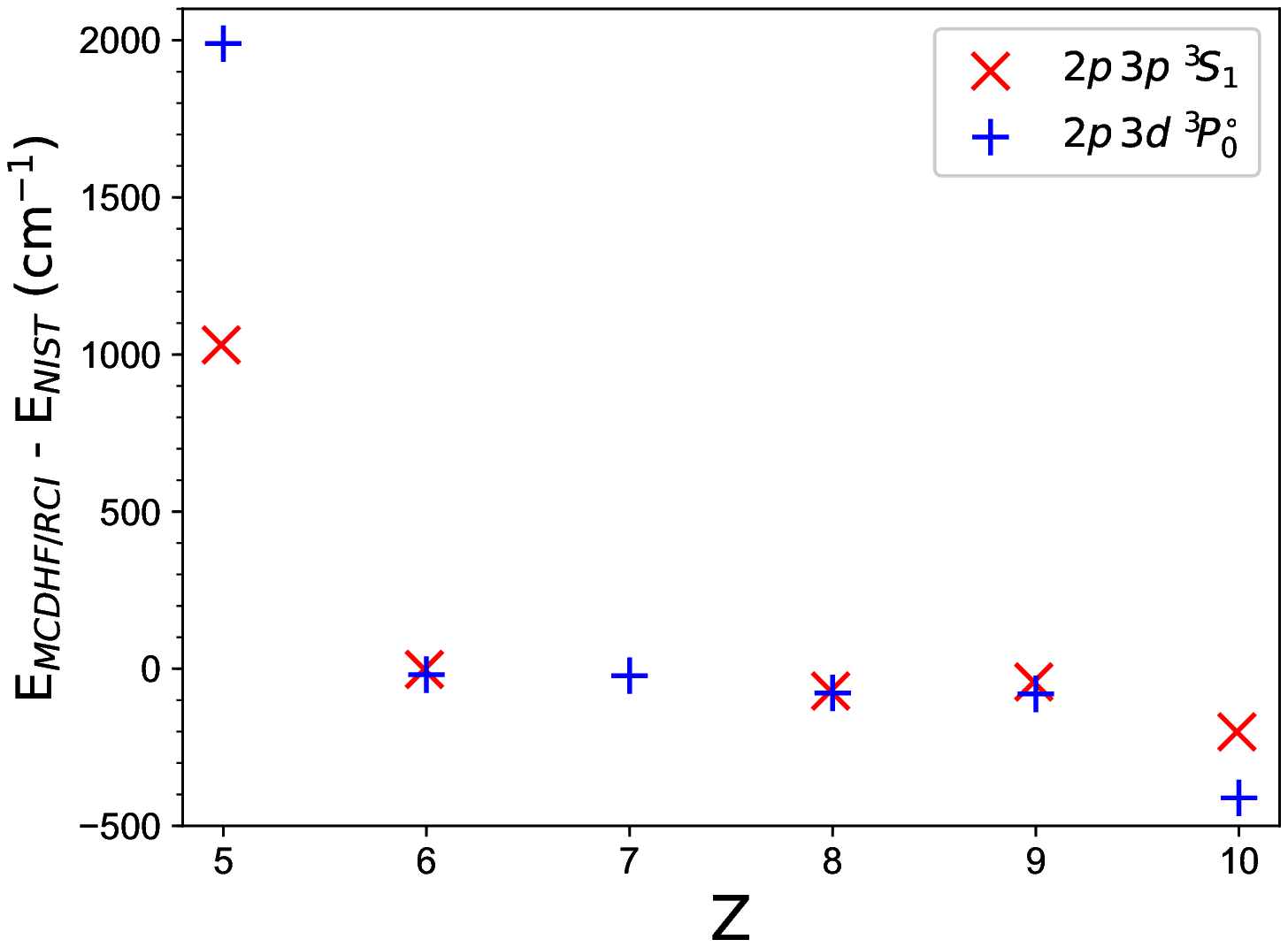}
	\caption{Energy differences in cm$^{-1}$ of the MCDHF/RCI and NIST excitation energies as a function of $Z$ for the $2p3p~^{3}S_{1}$ and $2p3d~^{3}P_{0}^{\circ}$ levels. The data come from Table~\ref{table.en.n23}.\label{figure.lev.nistlargedifferences}}
\end{figure*}

Among the levels $2p3p~^{3}S_{1}$ at $Z=5,7$, $2p3d~^{3}P_{0}^{\circ}$ at  $Z=5$, and  $2p3p~^{1}S_{0}$ at $Z=9$, where the differences of the MCDHF/RCI and NIST values are greater than 1000 cm$^{-1}$, the largest difference is -10451.5  cm$^{-1}$ for the level $\#29/2p3p~^{3}S_{1}$ at $Z=7$. The NIST energy 498045.5 cm$^{-1}$ for this level is in good agreement with our MCDHF/RCI value 498010 cm$^{-1}$ for the level $\#37/2s4s~^{3}S_{1}$. From Table~\ref{table.en} we can see that the weights for the first two eigenvector components of  the level $\#29$ are 70~\% ($2p3p~^{3}S_{1}$) and  27~\% ($2s4s~^{3}S$), whereas those of the level $\#37$ are 72~\% ($2s4s~^{3}S$) and  27~\% ($2p3p~^{3}S_{1}$). The two levels with relatively pure $LS$ coupling are mixed with each other, but the mixing is not strong. This implies that the NIST value 498045.5 cm$^{-1}$  should be designated 
as the $2s4s~^{3}S_{1}$ level. Apart from $2p3p~^{3}S_{1}$ at $Z=7$, we  cannot find obvious alternate designations for the other three levels in the present MCDHF/RCI calculations. As an example, the energy differences for the $2p3p~^{3}S_{1}$ and $2p3d~^{3}P_{0}^{\circ}$ levels as a function of $Z$ are shown in Figure~\ref{figure.lev.nistlargedifferences}.  Two anomalies exist for the $2p3p~^{3}S_{1}$ (the difference is 1030.5 cm$^{-1}$) and $2p3d~^{3}P_{0}^{\circ}$ (1989.4 cm$^{-1}$ ) levels at $Z=5$. Because the same scheme is used for each ion in the present MCDHF/RCI calculations, ensuring that the accuracy of our values for the same level is systematic and consistent, the large differences for these states are caused by relatively large uncertainty of the NIST values.

Lastly we compare the MCDHF/RCI excitation energies for the $n=4-6$ levels from \ion{B}{2} to \ion{Ne}{7}, which are not computed by~\citet[MCHF/BP]{Tachiev.1999.V32.p5805,Fischer.2004.V87.p1} nor by ~\citet[MBPT3 ]{Safronova.1996.V53.p4036,Safronova.1997.V30.p2375}, with the experimental values from the NIST ASD. Removing 3 obvious outliers, $\Delta E $ =  10~395.6, 2~659, and 6~258 cm$^{-1}$  for the levels $\#37/2s4s~^{3}S_{1}$ with $Z=7$, and $\#81/2p4s~^{3}P_{2}^{\circ}$  and $\#134/2p4p~^{1}S_{0}$  with $Z=8$, respectively, we find that for the remaining $n=4-6$ levels in Table~\ref{table.en}  $\Delta E \pm \sigma = -80 \pm 132$ cm$^{-1}$, which is highly satisfactory considering the range in excitation energy (exceeding \mbox{1 500 000} cm$^{-1}$).

\subsection{Transition rates}
Table~\ref{tab.tr.sub} lists transition rates $A$ of the AS$_5$ active set for the E1, M1, E2, and M2
transitions connecting all the levels included in Table~\ref{table.en}. To reduce the amount of data, only $A$ values greater than  $10^{-5}$ times the sum of $A$ values for all transitions from the upper level, i.e., radiative branching ratio (BR) greater than $10^{-5}$. Also included in this table are wavelengths $\lambda$, line strengths $S$, weighted oscillator strengths $gf$, and BR. The general rule is that transition data for electric multipoles (E1,E2) calculated in the Babushkin (length) gauge are preferable over transition data
calculated in the Coulomb (velocity) gauge. Whereas generally true, recent work~\citep{wang.2018,PehlivanRhodin.2017.V598.p102} has shown that computed transition data
from high states in neutral or near-neutral systems are more stable and show better convergence properties with respect to the increasing active set
of orbitals in the Coulomb gauge. For this reason, in Table~\ref{tab.tr.sub}, transition data from the $n=5,6$ states in B II and C III are computed in the Coulomb gauge. Transitions from all other states are computed in the Babushkin gauge.


Since line strengths $S$ for the transitions among the lowest 20 levels up to the $2s 3d$ configuration are provided by~\citet{Tachiev.1999.V32.p5805,Fischer.2004.V87.p1}~[MCHF/BP],  available in the NIST ASD~\citep{Kramida.2015.V.p}, in Table~\ref{table.tr.ne} we compare these data with the values  of the present MCDHF/RCI calculations using the AS$_5$ active set in \ion{Ne}{7}. Our MCDHF/RCI values of the AS$_4$ active set are also included in the table for comparison. Our two data sets (AS$_4$ and AS$_5$) agree  well within 3\%, which indicates that the present MCDHF/RCI calculations using the AS$_5$ active set are well converged. The MCHF/BP  values also show good agreement with our results except for the weak $2s3s~^{3}S_{1} - 2s2p~^{1}P_{1}^{\circ}$ and $2s3d~^{3}D_{1} - 2s2p~^{1}P_{1}^{\circ}$ intercombination transitions, which differ by more than 20\%.
It is also seen that transition data listed in the NIST ASD agree well with our results for most transitions. Differences larger that 20\% occur for transitions of relatively small radiative branching ratios only, the transition $2s3p~^{3}P_{1}^{\circ} - 2s^{2}~^{1}S_{0}$ being an exception. Although the $S$ value for this transition  is about $5 \times 10^5$, which is relatively small, the radiative branching ratio of this transition is 40.3\%. This means that the above transition gives an important contribution to the lifetime of the level $2s3p~^{3}P_{1}^{\circ}$. An inspection of Table~\ref{table.tr.ne} reveals that the accuracy of the NIST values is overestimated for some transitions, especially for the transitions with the estimated accuracy A ($\leq 3\%$), and the present MCDHF/RCI data should be  more accurate than those listed in the NIST ASD.

\begin{figure*}[ht!]
	\plotone{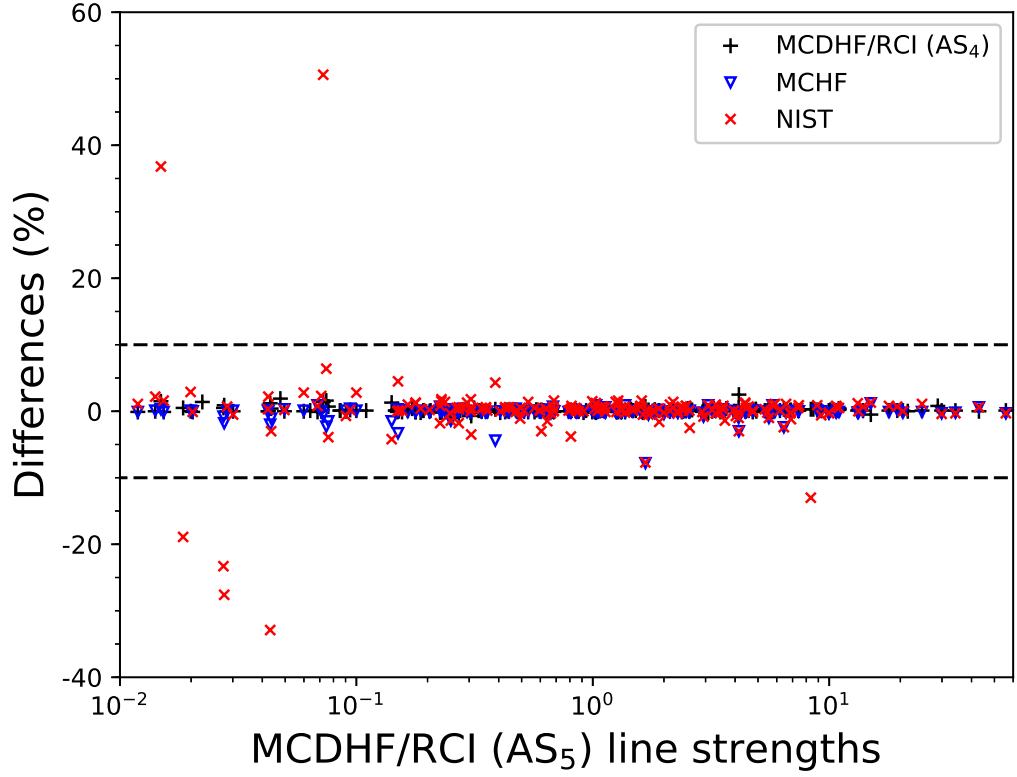}
	\caption{The differences ($S_{\rm other}/S_{\rm MCDHF/RCI (AS_5)}-1$) in \% of line strengths $S$ (in length form) for the transitions among the lowest 20 levels up to the $2s 3d$ configuration for the ions from \ion{B}{2} to \ion{Ne}{7} from different resources, MCDHF/RCI (AS$_4$), MCHF/BP, and NIST, relative to the values of the present MCDHF/RCI (AS$_5$) calculations. \label{fig.tr.n2-3}}
\end{figure*}

To further access the accuracy of the present MCDHF/RCI transition data, in Figure~\ref{fig.tr.n2-3}, we compare the line strengths $S$ from different resources (MCDHF/RCI (AS$_4$), MCHF/BP, and NIST) with the results of the present MCDHF/RCI (AS$_5$) calculations for the strong transitions with $S \geq 10^{-2}$ among the lowest 20 levels up to the $2s 3d$ configuration for the ions from \ion{B}{2} to \ion{Ne}{7}. Our two data sets (AS$_4$ and AS$_5$) agree very well within 5\% for all the transitions of the ions from \ion{B}{2} to \ion{Ne}{7}, which again indicates that the present MCDHF/RCI calculations using the AS$_5$ active set are well converged. The MCHF/BP  values also show good agreement with our results, within 10\%, whereas the NIST  values differ from the present calculations and the MCHF results by 10\% - 60\% for a few transitions. For example, the MCDHF/RCI (AS$_4$), MCDHF/RCI (AS$_5$), and MCHF/BP  values for the $2s3p~^{1}P_{1}^{\circ} - 2p^{2}~^{1}S_{0}$ transition are $7.34 \times 10^{-2}$, $7.24 \times 10^{-2}$, and $7.20 \times 10^{-2}$, respectively, while the corresponding NIST value is about 50\% greater ($1.09 \times 10^{-1}$). As a further check of the accuracy of the present transition data, we compare the $S$ values of our MCDHF/RCI (AS$_4$) and MCDHF/RCI (AS$_5$) calculations for transitions involving highly- excited states above the $2s 3d$ configuration  in \ion{Ne}{7}. 
Our two data sets agree within 10\% for most of the transitions. However, large deviations sometimes occur for weak one-photon-two-electrons transitions~\footnote{A one-photon two-electron transition is a transition between states of configurations that differ by two electrons, e.g. $2s^2~^1\!S_0 - 2p3d~^3\!P^o_1$. The rates of these transitions are strictly zero in the lowest-order approximation of the calculation and attain non-zero values as basis functions are expanded to describe electron correlation~\citep{Li.2010.V43.p35005}.}, which are generally very sensitive to  electron correlation effects and sometimes to high-order relativistic effects. 

\subsection{Lifetimes}
\begin{figure*}[ht!]
	\plotone{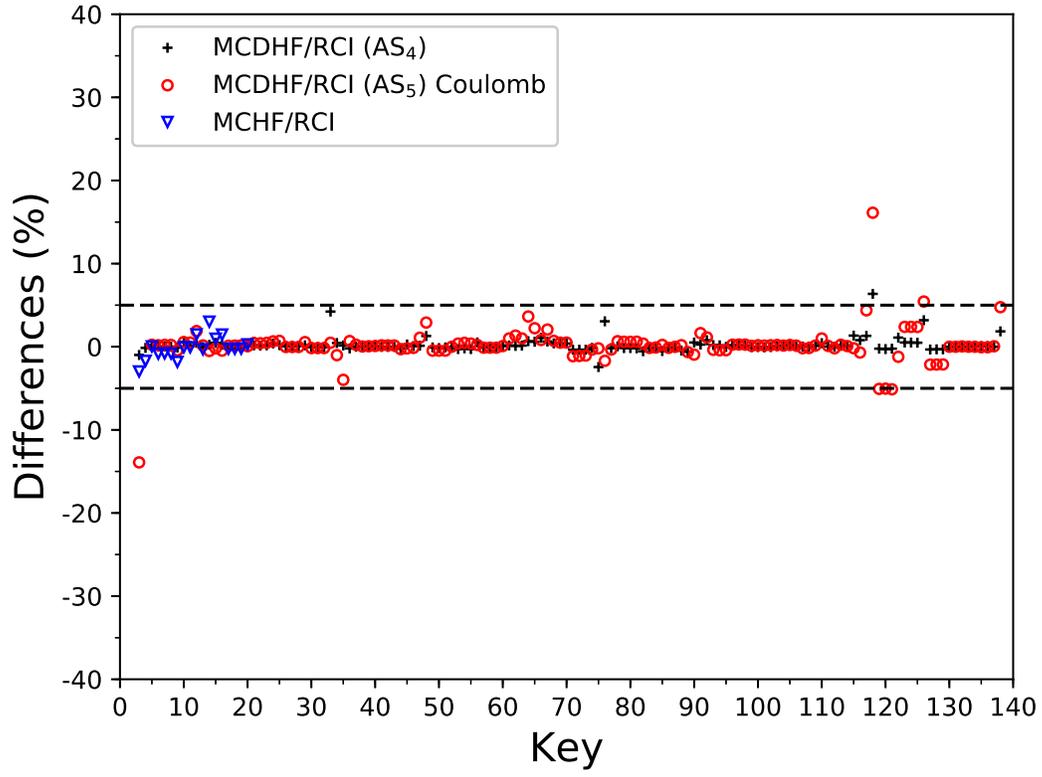}
	\caption{The differences ($\tau_{\rm other}/\tau_{\rm MCDHF/RCI (AS_5)}-1$) in \% of lifetimes $\tau$  for all the 138 levels in \ion{Ne}{7}  from different resources, MCDHF/RCI (AS$_4$) in Babushkin gauge, MCDHF/RCI (AS$_5$) in Coulomb gauge, and MCHF/BP, relative to the values of the present MCDHF/RCI (AS$_5$) calculations  in Babushkin gauge. \label{fig.lf.ne}}
\end{figure*}

Lifetimes in \ion{Ne}{7} obtained in the Babushkin gauge in the current work (MCDHF/RCI (AS$_4$) and reported by the MCHF/BP calculations ~\citep{Tachiev.1999.V32.p5805,Fischer.2004.V87.p1} are compared with the MCDHF/RCI (AS$_5$)) calculations in Figure~\ref{fig.lf.ne}. To further check the accuracy of the MCDHF/RCI lifetimes, we also calculate the lifetimes in the Coulomb gauge using the present AS$_5$ model. The results are also included in Figure~\ref{fig.lf.ne} for comparison. The agreement between the current lifetimes and  the values reported by ~\citet{Tachiev.1999.V32.p5805,Fischer.2004.V87.p1} is very good, well within 5\%. The differences of the present MCDHF/RCI (AS$_4$) and MCDHF/RCI (AS$_5$) calculations in the Babushkin gauge are within 5\% for almost all levels, with the largest deviation of 6\% for the level $2s6s~^{1}S_{0}$. The MCDHF/RCI (AS$_5$) values in the  Coulomb gauge also show good agreement (within 5\%) with the Babushkin-gauge values from the same calculation, with the two exceptions: $2s2p~^{3}P_{1}^{\circ}$ and $2s6s~^{1}S_{0}$. These two exceptions, with large deviations, occur for states with lifetimes determined by transitions with relatively small $S$ values. For example, the lifetime of the level $2s2p~^{3}P_{1}^{\circ}$ is determined by the weak intercombination transition $2s2p~^{3}P_{1}^{\circ}~-~2s^{2}~^{1}S_{0}$  with the $S$ value of $2 \times 10^{-5}$.  The contributions from the negative energy continuum, which are important for the transition rates in very weak intercombination transitions, are not included in the present MCDHF/RCI calculations in the Coulomb gauge. Therefore, the accuracy of the values in the Coulomb gauge, but not in the Babushkin gauge, are affected for the weak intercombination transition~\citep{Cheng.2008.V77.p52504}.

Lastly we compare our MCDHF/RCI lifetimes in the Babushkin gauge with the experimental values from different sources and the MCDHF/RCI~\citep{Joensson.1998.V31.p3497} and RCI~\citep{Cheng.2008.V77.p52504} calculations  for the $2s2p~^{1,3}P_{1}^{\circ}$ levels of the ions from \ion{B}{2} to \ion{Ne}{7}. The agreement between the current lifetimes and lifetimes reported by different experimental and theoretical sources is surprisingly good for both the resonance transition $2s2p~^{1}P_{1}^{\circ}$ and the intercombination transition $2s2p~^{3}P_{1}^{\circ}$. This is a strong indicator of the high accuracy of the present calculations.

\subsection{Summary}
Self-consistent MCDHF calculations and subsequent RCI calculations taking into account the Breit interaction and leading QED effects have been performed for the $n \leq 6$ states of Be-like ions from \ion{B}{2} to \ion{Ne}{7}. Level energies, transition energies, transition rates and lifetimes have been computed. Labeling of the states has been facilitated by converting from the $jj$-coupling to the $LSJ$-coupling scheme. Previous experimental and theoretical data involving low-lying $n \leq 3$ states along the isoelectronic sequence, as well as  both  low-lying and high-lying states for \ion{Ne}{7},  are used to validate the present computations. An excellent agreement with experimental data is found, which is manifested by an average energy difference with the standard deviation  for the $n \leq 3$ states of only $-23 \pm 141$ cm$^{-1}$ along the isoelectronic sequence. For all the 138 levels  of \ion{Ne}{7} the average  relative energy difference is $-0.01~\% \pm  0.02~\%$. Lifetimes in the Babushkin and Coulomb gauge agree to within 5\% for most states of \ion{Ne}{7}. We have presented calculations for energy and transition data that are more extensive and accurate than previous work.

To summarize, the present work has significantly increased the amount of accurate data for the Be-like isoelectronic sequence, extending our previous calculations~\citep{Wang.2015.V218.p16}.
The complete data set including both energy and transition results provided by the present work and by~\citet{Wang.2015.V218.p16}  can be used  for line identification and modeling purposes involving the $n > 3$ high-lying states, which fills the gap for lacking accurate atomic data on Be-like ions. Our study can also be considered as a benchmark for other calculations.

\section*{Scientific software packages}
Scientific software packages including~\software{GRASP2K \citep{Jonsson.2007.V177.p597,Jonsson.2013.V184.p2197}
and	FAC \citep{Gu.2008.V86.p675}} are used in the present work. 	

\acknowledgments
We thank the anonymous referee for many constructive comments on this manuscript. We acknowledge the support from the National Key Research and Development Program of China under Grant No. 2017YFA0403200, the National Natural Science Foundation of China (Grant Grant No. 11703004, No. 11674066, No.~11504421, and No.~11474034) and the Natural Science Foundation of Hebei Province, China (A2017201165). This work is also supported by the Swedish research council under contracts 2015-04842 and 2016-04185. 
The authors (K.W. and M.F.G.) express their gratitude to the support from the visiting researcher program at the Fudan University.

\clearpage
\bibliographystyle{aasjournal}
\bibliography{ref.bib}

\clearpage
\startlongtable
\begin{deluxetable*}{rrlrrrl}
	\tabletypesize{\scriptsize}
	\setlength{\tabcolsep}{4pt}
	\tablecaption{Excitation energies (in cm$^{-1}$)  and lifetimes ($\tau$ in s) for all the levels of Be-like ions with $Z=5-10$ from the present MCDHF/RCI calculations. The total angular momentum and pairty $J^{\pi}$ are also included in column 4. The rightmost column gives the $LS$ coupling expansion coefficients of the levels, in which the first number is the expansion coefficient  for the leading $LS$-composition.\label{table.en}}
	\tablehead{$Z$ & Key & Level& $J^{\pi}$& $E_{\rm MCDHF/RCI}$ & $\tau_{\rm{MCDHF/RCI}}$ & $LS$-composition}
	\startdata
10&    1& $2s^{2}~^{1}S_{0}$         &$0^e$ &         0  & 0.000E+00&       0.94 + 0.06~$2p^{2}~^{1}S$                                                                     \\
10&    2& $2s2p~^{3}P_{0}^{\circ}$   &$0^o$ &    111197  & 0.000E+00&       1.00                                                                                                              \\
10&    3& $2s2p~^{3}P_{1}^{\circ}$   &$1^o$ &    111649  & 5.277E-05&       1.00                                                                                                              \\
10&    4& $2s2p~^{3}P_{2}^{\circ}$   &$2^o$ &    112643  & 1.419E+01&       1.00                                                                                                              \\
10&    5& $2s2p~^{1}P_{1}^{\circ}$   &$1^o$ &    215065  & 2.508E-10&       0.99                                                                                                              \\
10&    6& $2p^{2}~^{3}P_{0}$         &$0^e$ &    289259  & 3.249E-10&       0.99                                                                                                              \\
10&    7& $2p^{2}~^{3}P_{1}$         &$1^e$ &    289771  & 3.235E-10&       0.99                                                                                                              \\
10&    8& $2p^{2}~^{3}P_{2}$         &$2^e$ &    290653  & 3.213E-10&       0.99                                                                                                              \\
\enddata
	\tablecomments{ Table~\ref{table.en} is published in its entirety in the machine-readable format. The values for \ion{Ne}{7} are shown here for guidance regarding its form and content.}
\end{deluxetable*}

\newpage

\startlongtable
	\begin{deluxetable*}{rlrrrrrrrr}
		\tablecaption{The MCDHF/RCI excitation energies in cm$^{-1}$ for \ion{Ne}{7} as a function of the increasing size of the CSF expansion. AS$_1$,  AS$_2$, AS$_3$, AS$_4$, and AS$_5$ -- the present MCDHF/RCI excitation energies; NIST -- the compiled values from the NIST ASD~\citep{Kramida.2015.V.p}; $\Delta E$ -- the differences between the MCDHF/RCI calculated energies for AS$_5$ and the NIST compiled values.\label{table.as}}
		\tablewidth{700pt}
		\tabletypesize{\scriptsize}
		\tablehead{
			\colhead{Key} & \colhead{Level}& \multicolumn{5}{c}{ MCDHF/RCI}&\colhead{NIST}&\colhead{$\Delta E$}
 \\ 
 \cmidrule(r){3-7}
 	\colhead{}&	\colhead{}&  			\colhead{AS$_1$} & \colhead{AS$_2$} & 
 \colhead{AS$_3$} & \colhead{AS$_4$} & 
 \colhead{AS$_5$} 
		} 
\startdata
1   & $2s^{2}~^{1}S_{0}$          &      0&	      0&	      0&	      0&	      0&       0      &  \nodata     \\                          
2   & $2s2p~^{3}P_{0}^{\circ}$    & 111669&	 111132&	 111139&	 111182&	 111197&  111252.8    & -55.8        \\
3   & $2s2p~^{3}P_{1}^{\circ}$    & 112120&	 111585&	 111591&	 111635&	 111649&  111708      & -59          \\
4   & $2s2p~^{3}P_{2}^{\circ}$    & 113113&	 112578&	 112585&	 112628&	 112643&  112702.4    & -59.4        \\
5   & $2s2p~^{1}P_{1}^{\circ}$    & 221483&	 216852&	 215442&	 215156&	 215065&  214951.6    & 113.4        \\
6   & $2p^{2}~^{3}P_{0}$          & 291107&	 289527&	 289213&	 289244&	 289259&  289330.4    & -71.4        \\
7   & $2p^{2}~^{3}P_{1}$          & 291616&	 290038&	 289724&	 289756&	 289771&  289841.1    & -70.1        \\
8   & $2p^{2}~^{3}P_{2}$          & 292492&	 290920&	 290606&	 290638&	 290653&  290724      & -71          \\
\enddata
\end{deluxetable*}

\cleardoublepage

\startlongtable
\begin{deluxetable*}{rlrrrrrrrrrrrrr}
	\tablecaption{Computed excitation energies of \ion{Ne}{7} from the present MCDHF/RCI and MBPT calculations and from the calculations by~\citet{Savukov.2006.V39.p2115} [MBPT2], by ~\cite{Menchero.2014.V566.p104} [AUTOSTRUCTURE],  by ~\citet{Safronova.1996.V53.p4036,Safronova.1997.V30.p2375} [MBPT3], and  by~\citet{Tachiev.1999.V32.p5805,Fischer.2004.V87.p1} [MCHF/BP] are compared with the NIST compiled values. $E$ -- excitation energies from different resources ; $\Delta E$ -- the differences between the calculated energies from different resources and the NIST compiled values.\label{table.en.ne}}
	\tablewidth{500pt}
	\tabletypesize{\tiny}
	\tablehead{
		\colhead{}&	\colhead{}&  			\colhead{NIST} & \multicolumn{2}{c}{ MCDHF/RCI} & 
\multicolumn{2}{c}{ MBPT} & \multicolumn{2}{c}{ MBPT2} & \multicolumn{2}{c}{AUTOSTRUCTURE}  & \multicolumn{2}{c}{ MBPT3} &
\multicolumn{2}{c}{MCHF/BP} 
\\
\cmidrule(r){3-3} \cmidrule(r){4-5}\cmidrule(r){6-7} \cmidrule(r){8-9}\cmidrule(r){10-11}\cmidrule(r){12-13}\cmidrule(r){14-15}
		\colhead{Key} & \colhead{Level}&\colhead{$E$}&\colhead{$E$}&\colhead{$\Delta E$}&\colhead{$E$}&\colhead{$\Delta E$}&\colhead{$E$}&\colhead{$\Delta E$}&\colhead{$E$}&\colhead{$\Delta E$}&\colhead{$E$}&\colhead{$\Delta E$}&\colhead{$E$}&\colhead{$\Delta E$}
	} 
	\startdata
1   & $2s^{2}~^{1}S_{0}$         &      0      &0       &\nodata& 0       &\nodata & 0       & \nodata &0      &\nodata    & 0       &\nodata  & 0       &\nodata \\                          
2   & $2s2p~^{3}P_{0}^{\circ}$   & 111252.8    &111197  &-55.8  & 111182  & -70.8  & 111264  & 11.2    &111924 &671.2      & 111128  & -124.8  & 111441  & 188.2  \\
3   & $2s2p~^{3}P_{1}^{\circ}$   & 111708      &111649  &-59    & 111592  & -116   & 111708  & 0       &112461 &753        & 111582  & -126    & 111896  & 188    \\
4   & $2s2p~^{3}P_{2}^{\circ}$   & 112702.4    &112643  &-59.4  & 112640  & -62.4  & 112706  & 3.6     &113551 &848.6      & 112578  & -124.4  & 112893  & 190.6  \\
5   & $2s2p~^{1}P_{1}^{\circ}$   & 214951.6    &215065  &113.4  & 214844  & -107.6 & 215524  & 572.4   &222371 &7419.4     & 213581  & -1370.6 & 215260  & 308.4  \\
6   & $2p^{2}~^{3}P_{0}$         & 289330.4    &289259  &-71.4  & 289073  & -257.4 & 289683  & 352.6   &293008 &3677.6     & 289144  & -186.4  & 290029  & 698.6  \\
7   & $2p^{2}~^{3}P_{1}$         & 289841.1    &289771  &-70.1  & 289606  & -235.1 & 290190  & 348.9   &293562 &3720.9     & 289657  & -184.1  & 290546  & 704.9  \\
8   & $2p^{2}~^{3}P_{2}$         & 290724      &290653  &-71    & 290466  & -258   & 291064  & 340     &294617 &3893       & 290544  & -180    & 291429  & 705    \\
\enddata
\end{deluxetable*}

\cleardoublepage

\startlongtable
\begin{deluxetable*}{rrlrrrrrrrrrrr}
	\tablecaption{Computed excitation energies for the $n \leq 3$ levels of the ions from \ion{B}{2} to \ion{Ne}{7} from the present MCDHF/RCI calculations, and from the calculations by ~\citet{Safronova.1996.V53.p4036,Safronova.1997.V30.p2375} [MBPT3], and  by~\citet{Tachiev.1999.V32.p5805,Fischer.2004.V87.p1} [MCHF/BP] are compared with the NIST compiled values. $E$ -- excitation energies from different resources ; $\Delta E$ -- the differences between the calculated energies from different resources and the NIST compiled values.\label{table.en.n23}}
	\tablewidth{500pt}
	\tabletypesize{\scriptsize}
	\tablehead{
		\colhead{}&	\colhead{}&\colhead{}&  			\colhead{NIST} & \multicolumn{2}{c}{ MCDHF/RCI} & 
		 \multicolumn{2}{c}{ MBPT3} &
		\multicolumn{2}{c}{MCHF/BP} 
		\\
		\cmidrule(r){4-4} \cmidrule(r){5-6}\cmidrule(r){7-8} \cmidrule(r){9-10}
		\colhead{Z}&\colhead{Key} & \colhead{Level}&\colhead{$E$}&\colhead{$E$}&\colhead{$\Delta E$}&\colhead{$E$}&\colhead{$\Delta E$}&\colhead{$E$}&\colhead{$\Delta E$}
	} 
	\startdata
5 &1    & $2s^{2}~^{1}S_{0}$                  & 0              &0       &\nodata  &0              &\nodata        &0              &\nodata              \\
5 &2    & $2s2p~^{3}P_{0}^{\circ}$    & 37335.54       &37326   &-9.54    &36624          &-711.54        &37468          &132.46               \\
5 &3    & $2s2p~^{3}P_{1}^{\circ}$    & 37341.65       &37331   &-10.65   &36631          &-710.65        &37475          &133.35               \\
5 &4    & $2s2p~^{3}P_{2}^{\circ}$    & 37357.8        &37347   &-10.8    &36648          &-709.8         &37491          &133.2                \\
5 &5    & $2s2p~^{1}P_{1}^{\circ}$    & 73396.51       &73670   &273.49   &70046          &-3350.51       &73583          &186.49               \\
5 &6    & $2p^{2}~^{3}P_{0}$       & 98911.38       &98911   &-0.38    &98042          &-869.38        &99236          &324.62               \\
5 &7    & $2p^{2}~^{3}P_{1}$       & 98919.87       &98919   &-0.87    &98051          &-868.87        &99245          &325.13               \\
5 &8    & $2p^{2}~^{3}P_{2}$       & 98933.26       &98932   &-1.26    &98066          &-867.26        &99258          &324.74               \\
\enddata
\end{deluxetable*}

\cleardoublepage

\startlongtable
\begin{deluxetable*}{rrrcccccccccl}
	\tabletypesize{\scriptsize}
	\tablecolumns{7}
	\tablewidth{0pt}
	\tablecaption{{\textnormal {Transition data for Be-like ions from \ion{B}{2} to \ion{Ne}{7} from the present MCDHF/RCI calculations. $\lambda$ is transition wavelength in vacuum in {\AA}, \emph{S} is line strength in atomic units, $gf$ is weighted oscillator strength, and \emph{A} is transition rate in s$^{-1}$. Radiative branching ratios (BR) for each transition provided by the present MCDHF/RCI calculations are listed in the last column. Only transitions contributing a fraction of more than $10^{-5}$ to the total radiative lifetime  are presented.}} \label{tab.tr.sub}}
	\tablehead{$Z$ & \emph{j} & \emph{i} &  $\lambda_{\rm MCDHF/RCI}$ &Type  & $\emph{S}_{\rm MCDHF/RCI}$ & \emph{gf}$_{\rm MCDHF/RCI}$  & $A_{\rm MCDHF/RCI}$ &BR \vspace{ -4mm} \\}
	\startdata
 10&   3&   1&   8.95657E+02&   E1& 2.016E-05&  6.837E-06&  1.895E+04&    1.000E+00\\  
10&   4&   1&   8.87759E+02&   M2& 1.059E+01&  3.384E-11&  5.728E-02&    8.126E-01\\  
10&   4&   3&   1.00669E+05&   M1& 2.499E+00&  1.004E-07&  1.321E-02&    1.874E-01\\  
10&   5&   1&   4.64974E+02&   E1& 5.935E-01&  3.877E-01&  3.988E+09&    1.000E+00\\  
10&   6&   3&   5.63031E+02&   E1& 2.711E-01&  1.463E-01&  3.078E+09&    1.000E+00\\  
10&   6&   5&   1.34781E+03&   E1& 3.803E-05&  8.572E-06&  3.147E+04&    1.023E-05\\  
10&   7&   2&   5.59992E+02&   E1& 2.714E-01&  1.472E-01&  1.044E+09&    3.377E-01\\  
10&   7&   3&   5.61414E+02&   E1& 2.035E-01&  1.101E-01&  7.766E+08&    2.512E-01\\
\enddata
	\tablenotetext{\space}{Table~\ref{tab.tr.sub} is published in its entirety in the machine-readable format. A portion is shown here for guidance regarding its form and content.}
\end{deluxetable*}

\cleardoublepage

\startlongtable
\begin{deluxetable*}{llccccccccc}
\tablecaption{The differences in \% of line strengths $S$ (in length form) for the transitions among the lowest 20 levels up to the $2s 3d$ configuration of \ion{Ne}{7} from different resources, MCDHF/RCI (AS$_4$), MCHF/BP, and NIST, relative to the values of the present MCDHF/RCI (AS$_5$) calculations. Radiative branching ratios (BR) for each transition from the present MCDHF/RCI (AS$_5$) calculations are listed in the last column.  MCDHF/RCI (AS$_4$) - the present MCDHF/RCI values of the AS$_4$ active set; MCDHF/RCI (AS$_5$) - the present MCDHF/RCI values of the AS$_5$ active set; MCHF/BP - the values provided by \citet{Tachiev.1999.V32.p5805,Fischer.2004.V87.p1}; NIST -the values included in the NIST ASD, as well as the estimated accuracies of transition data using the terminology of the NIST ASD~\citep{Kramida.2015.V.p}.\label{table.tr.ne}}
\tablewidth{500pt}
\tabletypesize{\scriptsize}
\tablehead{
\colhead{Upper level}& \colhead{Lower level}& \colhead{Type}& \multicolumn{2}{c}{ MCDHF/RCI}& \colhead{MCHF/BP}& \colhead{NIST}& \multicolumn{3}{c}{Differences (\%)}& \colhead{BR}\\
\cmidrule(r){4-5}  \cmidrule(r){8-10}
&& &\colhead{AS$_4$}&\colhead{AS$_5$}&&&\colhead{AS$_4$}&\colhead{MCHF/BP}&\colhead{NIST}
	} 
	\startdata
$2s2p~^{3}P_{1}^{\circ}$	&$2s^{2}~^{1}S_{0}$              &     E1&  2.01E-05& 2.02E-05& 2.04E-05& 1.90E-05$^{\rm C}$&-0.3     &1.0      &-5.8     &1.00E+00 \\ 
$2s2p~^{3}P_{2}^{\circ}$	&$2s^{2}~^{1}S_{0}$              &     M2&  1.06E+01& 1.06E+01& 1.06E+01& 1.06E+01$^{\rm A}$&0.0      &-0.2     &0.1      &8.13E-01 \\ 
$2s2p~^{3}P_{2}^{\circ}$	&$2s2p~^{3}P_{1}^{\circ}$        &     M1&  2.50E+00& 2.50E+00& 2.50E+00& 2.50E+00$^{\rm B}$&0.0      &0.0      &0.0      &1.87E-01 \\ 
$2s2p~^{1}P_{1}^{\circ}$	&$2s^{2}~^{1}S_{0}$              &     E1&  5.93E-01& 5.94E-01& 5.94E-01& 5.86E-01$^{\rm A}$&-0.1     &0.0      &-1.3     &1.00E+00 \\ 
$2p^{2}~^{3}P_{0}$   	    &$2s2p~^{3}P_{1}^{\circ}$        &     E1&  2.71E-01& 2.71E-01& 2.71E-01& 2.72E-01$^{\rm A}$&0.0      &0.0      &0.3      &1.00E+00 \\ 
$2p^{2}~^{3}P_{0}$   	    &$2s2p~^{1}P_{1}^{\circ}$        &     E1&  3.78E-05& 3.80E-05& 3.87E-05& 3.80E-05$^{\rm D}$&-0.5     &1.7      &-0.1     &1.02E-05 \\ 
$2p^{2}~^{3}P_{1}$   	    &$2s2p~^{3}P_{0}^{\circ}$        &     E1&  2.71E-01& 2.71E-01& 2.71E-01& 2.73E-01$^{\rm A}$&0.0      &-0.1     &0.6      &3.38E-01 \\ 
$2p^{2}~^{3}P_{1}$   	    &$2s2p~^{3}P_{1}^{\circ}$        &     E1&  2.04E-01& 2.04E-01& 2.03E-01& 2.04E-01$^{\rm A}$&0.0      &-0.1     &0.2      &2.51E-01 \\
\enddata
\end{deluxetable*}

\begin{table*}
\begin{center}
	\caption{Comparisons between the experimental and theoretical lifetimes (in s) for the $2s2p~^{1,3}P_{1}^{\circ}$ levels of the ions from \ion{B}{2} to \ion{Ne}{7}.\label{table.lf}}
	\begin{tabular}{lllrrllll}
		\tableline\tableline
		\colhead{\emph{Z}} &\colhead{Key}  &\colhead{State}  &\colhead{Exp.\tablenotemark{a}} &\colhead{Uncertainty\tablenotemark{a}}
		&\colhead{MCDHF/RCI\tablenotemark{b}}  &\colhead{MCDHF/RCI\tablenotemark{c}}  &\colhead{RCI\tablenotemark{d}}\\
		\tableline
		5&  3& $2s2p~^{3}P_{1}^{\circ}$& 9.765E-02\tablenotemark{[e]} & 0.05E-02   &  9.513E-02 & \nodata   & \nodata \\
6&  3& $2s2p~^{3}P_{1}^{\circ}$& 9.714E-03\tablenotemark{[f]} & 0.013E-03 &  9.795E-03 & \nodata   & 9.970E-03\\
7&  3& $2s2p~^{3}P_{1}^{\circ}$& 1.73E-03 \tablenotemark{[g]} & 0.01E-03  &  1.751E-03 & 1.790E-03 & 1.766E-03\\
8&  3& $2s2p~^{3}P_{1}^{\circ}$& 4.32E-04 \tablenotemark{[g]} & 0.09E-04  &  4.440E-04 & 4.521E-04 & 4.462E-04\\
9&  3& $2s2p~^{3}P_{1}^{\circ}$& \nodata                      & \nodata   &  1.437E-04 & 1.418E-04\\
10&  3& $2s2p~^{3}P_{1}^{\circ}$& 4.5E-05  \tablenotemark{[h]} & 2.3E-05   &  5.277E-05 & 5.371E-05 & 5.299E-05\\
\\                                                           
5&  5& $2s2p~^{1}P_{1}^{\circ}$& 8.6E-10  \tablenotemark{[i]}& 0.7E-10 &  8.285E-10 & \nodata   & \nodata    \\
6&  5& $2s2p~^{1}P_{1}^{\circ}$& 5.7E-10  \tablenotemark{[j]}& 0.2E-10 &  5.634E-10 & \nodata   & 5.672E-10   \\
7&  5& $2s2p~^{1}P_{1}^{\circ}$&4.25E-10  \tablenotemark{[k]}& 0.15E-10&  4.299E-10 & 4.285E-10 & 4.316E-10   \\
8&  5& $2s2p~^{1}P_{1}^{\circ}$&3.38E-10  \tablenotemark{[k]}& 0.15E-10&  3.480E-10 & 3.466E-10 & 3.490E-10   \\
9&  5& $2s2p~^{1}P_{1}^{\circ}$& 3.1E-10  \tablenotemark{[l]}& 0.4E-10 &  2.919E-10 & 2.907E-10 & 2.925E-10   \\
10&  5& $2s2p~^{1}P_{1}^{\circ}$&2.32E-10  \tablenotemark{[m]}&0.37E-10 &  2.508E-10 & 2.498E-10 & 2.511E-10   \\
		\tableline
	\end{tabular}
\end{center}
	\tablenotetext{a}{Experimental lifetimes and experimental uncertainties.}
	\tablenotetext{b}{The present MCDHF/RCI lifetimes.}
	\tablenotetext{c}{The MCDHF/RCI lifetimes calculated by \citet{Joensson.1998.V31.p3497}.}
	\tablenotetext{d}{The RCI lifetimes calculated by \citet{Cheng.2008.V77.p52504}.}
	\tablenotetext{}{The experimental lifetimes from \citet{Traebert.1999.V32.p537}$^{[\rm e]}$, \citet{Doerfert.1997.V78.p4355}$^{[\rm f]}$, \citet{Traebert.2005.V38.p2395}$^{[\rm g]}$, \citet{Kunze.1972.V13.p565} $^{[\rm h]}$, \citet{Bashkin.1985.V9.p593}$^{[\rm i]}$,  \citet{Reistad.1986.V34.p151}$^{[\rm j]}$, \citet{Engstroem.1981.V24.p551}$^{[\rm k]}$, \citet{	Knystautas.1979.V69.p474}$^{[\rm l]}$, and \citet{Irwin.1973.V51.p1948}$^{[\rm m]}$.}
\end{table*}
\listofchanges
\end{document}